\newcommand{\bm}[1]{\mbox{\boldmath$#1$}}

\documentclass[twocolumn,showpacs,preprintnumbers,amsmath,amssymb]{revtex4}

\usepackage{graphicx}
\usepackage{dcolumn}
\usepackage{bm}

\begin{document}

\title{Reentrant spin-glass transition in a dilute magnet }

\author{ S. Abiko} 
\author{ S. Niidera} 
\author{ F. Matsubara} 

\affiliation{Department of Applied Physics, Tohoku University, Sendai 980-8579,
Japan\\}

\date{ \today }

\begin{abstract}

We have performed a large scale Monte Carlo simulation of a dilute 
classical Heisenberg model with ferromagnetic nearest neighbor 
and antiferromagnetic next-nearest neighbor interactions. 
We found that {\it the model reproduces a reentrant spin-glass 
transition}. 
That is, as the temperature is decreased, the magnetization
increases rapidly below a certain temperature, reaches a maximum value, 
then ceases at some lower temperature. 
The low temperature phase was suggested to be a spin-glass phase 
that is characterized by ferromagnetic clusters. 

\end{abstract}


\maketitle


Reentrant spin-glass (RSG) transition is a well-known phenomenon of 
spin glasses (SGs). 
The RSG transition is found near the phase boundary between the SG phase 
and the ferromagnetic phase\cite{Verbeek,BinderY}. 
As the temperature decreases from a higher temperature, 
magnetization increases, then ceases at a lower temperature. 
Finally, the SG phase is realized. 
The phenomenon was first considered as a phase transition between 
a `ferromagnetic phase' and a `SG phase'\cite{Yeshurun}. 
However, neutron diffraction studies have revealed that the `SG phase' 
is characterized by ferromagnetic clusters\cite{Maletta,Aeppli,Motoya}. 
Now the RSG transition is believed to be a re-entry from a ferromagnetic 
phase to a frozen state with ferromagnetic clusters.

The mechanism responsible for this reentrant transition has 
not yet been resolved. 
Two ideas have been proposed for describing the RSG transition: 
(i) an infinite-range Ising bond model, and (ii) a phenomenological 
random field concept.
Sherrington and Kirkpatrick solved the infinite-range Ising bond model 
using a replica technique\cite{SKmodel}. 
They predicted the occurrence of the RSG phase transition 
before the RSG transition was observed experimentally\cite{Verbeek}. 
However, successive studies of the model revealed that the RSG phase 
transition does not occur\cite{Toulouse}, even when the vector 
spins substitute for the Ising spins\cite{GT}.
On the other hand, the random field idea was proposed to explain 
experimental observations of neutron scattering 
functions\cite{Maletta,Aeppli,Motoya,Arai}. 
The essential point of that conception is that the system is decomposed into 
a ferromagnetic part (FM part) and a part with frustrated spins (SG part). 
At low temperatures, the spins of the SG part will yield random effective 
fields to the spins of the FM part. The FM order, which grows at higher 
temperatures, vanishes because of a random field effect\cite{ImryMa}. 
Later, it was pointed out that not random field but random anisotropy 
brings a RSG-like phenomenon\cite{Chudnovsky,Fisch}. 
In a FM model with random anisotropy, as the temperature is decreased from 
a high temperature, there first appears a correlated SG (CSG) phase 
that is characterized by a well developed FM spin correlation. 
Then the CSG phase will become transformed into a speromagnet, i.e., 
a random SG-like phase, just like the RSG transition\cite{Chudnovsky}. 
Nevertheless, no theoretical evidence has yet been presented on this idea in 
a microscopic point of view. 
In the last two decades, computer simulations have been performed extensively 
to solve the RSG transition in various models such as short-range bond 
models\cite{Ozeki,Iyota,Gingras}, short-range site models\cite{Binder,Tamiya} 
and a Ruderman-Kittel-Kasuya-Yoshida model\cite{Iguchi,Morishita}. 
However, no realistic model has been shown to reproduce 
vanishing magnetization that grows at higher temperatures.

This Letter reports a dilute classical Heisenberg model 
that reproduces a RSG transition. 
That is, the magnetization increases rapidly below a certain temperature, 
then disappears at some lower temperature. 
The low temperature phase was suggested to be a SG phase, which is 
characterized by ferromagnetic clusters. 
It should be emphasized that no novel mechanism is necessary to 
reproduce the RSG transition. 
The model studied here is a natural one that was proposed 
experimentally\cite{EuSrS} and investigated using 
a Monte Carlo (MC) method\cite{Binder}. 
A large-scale MC simulation revealed the nature of the model. 
We believe that the system size is crucial for vanishing the magnetization, 
because the SG phase at low temperatures is composed 
of ferromagnetic clusters. 


We start with a dilute Heisenberg model with competing nearest and 
next-nearest neighbor exchange interactions described by the Hamiltonian: 
\begin{eqnarray} 
 H = &-& \sum_{\langle ij \rangle}^{nn}J_1x_ix_j\bm{S}_{i}\cdot\bm{S}_{j} 
   + \sum_{\langle kl \rangle}^{nnn}J_2x_kx_l\bm{S}_{k}\cdot\bm{S}_{l}, 
\end{eqnarray} 
where $\bm{S}_{i}$ is the classical Heisenberg spin of $|\bm{S}_{i}| = 1$; 
$J_1 (> 0)$ and $J_2 (> 0)$ respectively represent the nearest neighbor 
and the next-nearest neighbor exchange interactions; $x_i = 1$ or 0 when the 
lattice site $i$ is occupied respectively by a magnetic or non-magnetic atom. 
The average number of $x (\equiv \langle x_i \rangle)$ is the concentration 
of a magnetic atom. 
Note that an experimental realization of this model is 
Eu$_x$Sr$_{1-x}$S\cite{EuSrS}, in which magnetic atoms (Eu) are located on 
the fcc lattice sites\cite{Comm_EuSrS}. 
Here, for simplicity, we consider the model on a simple cubic lattice with 
$J_2 = 0.2J_1$\cite{Model}.

A computer simulation was performed using a conventional heat bath MC method. 
The system was cooled gradually from a high temperature (cooling simulation). 
We calculated the magnetization $M$ defined as $M = [\langle M(s) \rangle]$, 
where $M(s) (\equiv |\sum_ix_i\bm{S}_i|)$ is the magnetization at the 
$s$th MC step, and  
%
%
$\langle \cdots \rangle$ represents a MC average 
and $[ \cdots ]$ a sample average. 
Here, for larger lattices, $200 000$ MC steps (MCS) were allowed for 
relaxation; data of successive $200 000$ MCS were used to calculate 
average values. 
We will show later that these MCS are sufficient for studying 
equilibrium properties of the model at a temperature range 
within which the RSG behavior is found. 
We made a simulation for $x > 0.70$. 
We treated lattices of $L \times L \times L \ (L= 8-48)$ with 
periodic boundary conditions. 
Numbers $N_s$ of samples with different spin distributions are 
$N_s = 1000$ for $L \leq 12$, $N_s = 600$ for $L = 16$, 
$N_s = 200$ for $L = 24$ and 32, and $N_s = 80$ for $L = 48$. 
We measured the temperature in the unit of $J_1$ ($k_{\rm B} = 1$).

\begin{figure}[tb]
\includegraphics[width=7.0cm,clip]{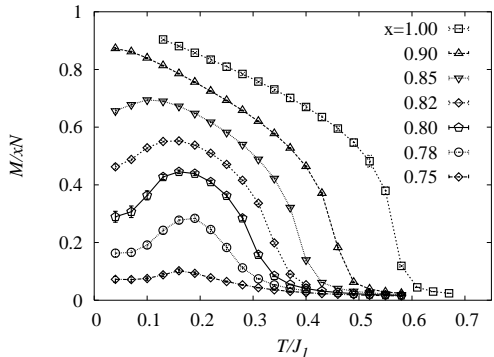}
\vspace{-0.4cm}
\caption{\label{fig:1}
Magnetizations $M$ in the $32 \times 32 \times 32$ lattice for 
various spin concentrations $x$. 
}
\end{figure}

Figure 1 shows temperature dependencies of the magnetization per spin $M/xN$ 
for various $x$, where $N (\equiv L^3)$ is the number of lattice sites. 
For $x = 1$, as the temperature decreases, $M$ increases rapidly 
below a temperature revealing the occurrence of a ferromagnetic phase. 
As $x$ decreases, $M$ exhibits an interesting behavior. 
In the range of $0.78 \lesssim x \lesssim 0.85$, $M$ once increases, 
reaches a maximum value, then decreases. 
Such behavior of $M$ is reminiscent of the occurrence of the RSG transition.

\begin{figure}[tb]
\includegraphics[width=7.0cm,clip]{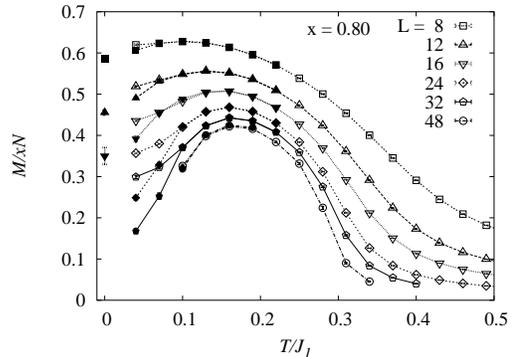}
\vspace{-0.4cm}
\caption{\label{fig:2}
Magnetizations $M$ for $x = 0.80$ in the $L\times L\times L$ lattice. 
Open symbols indicate $M$ in the cooling simulation and filled symbols 
that in the heating simulation. Data at $T = 0$ indicate those 
in the ground state. 
}
\end{figure}

To examine this phenomenon, we made detailed studies in the case of 
$x = 0.80$.
First we note that we performed a complementary simulation. 
That is, starting with a random spin configuration at a low temperature, 
the system is heated gradually (heating simulation). 
In addition, we investigated $M$ in the ground state for smaller lattices 
($L \leq 16$) having used a hybrid genetic algorithm\cite{GA}. 
Figure 2 shows temperature dependencies of $M$ in both cooling and 
heating simulations for various $L$ together with the value of $M$ 
in the ground state. 
For $T \gtrsim 0.1J_1$, data of the two simulations almost coincide 
mutually, even for large $L$. 
We thereby infer that $M$ for $T \gtrsim 0.1J_1$ are of thermal 
equilibrium. 
On the other hand, for $T < 0.1J_1$, a great difference in $M$ is seen 
between the two simulations; estimation of the equilibrium value is difficult. 
We speculate that the heating simulation gives a value of $M$ 
that is similar to that in the equilibrium state because the data 
in the heating simulation seem to connect to those in the ground state.

Figure 2 shows that the lattice size dependence of $M$ is remarkable. 
For smaller $L$, as the temperature decreases, $M$ decreases slightly 
at very low temperatures. The decrease is enhanced as $L$ increases. 
Consequently, a strong size-dependence of $M$ is indicated for 
$T \lesssim 0.1J_1$. 
Particularly in the ground state, $M$ apparently decreases rapidly as 
$L$ increases. 
These facts imply that $M$ disappears at low temperatures 
as well as at high temperatures.

\begin{figure}[tb]
\begin{center}
\includegraphics[width=7.5cm,clip]{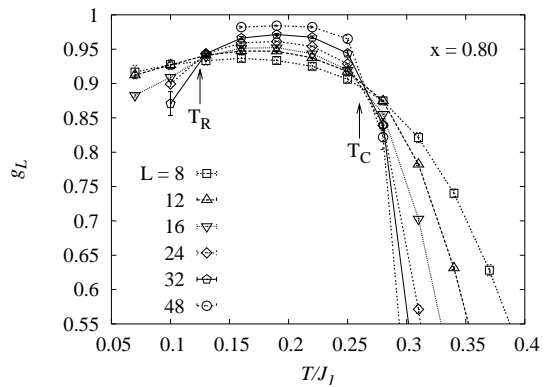}\\
\end{center}
\vspace{-0.4cm}
\caption{\label{fig:3}
Binder parameters $g_L$ for $x = 0.80$. 
}
\end{figure}

We investigated the Binder parameter $g_L$, 
\begin{eqnarray} 
 g_L = (5 - 3\frac{[\langle M(s)^4\rangle]}{[\langle M(s)^2\rangle]^2})/2, 
\end{eqnarray} 
to examine whether or not the ferromagnetic phase actually occurs within 
an intermediate temperature range. 
Figure 3 shows $g_L$'s for various $L$\cite{Comm_gL}. 
We can see that $g_L$'s for different $L$ cross at two 
temperatures $T_{\rm C}$ and $T_{\rm R}$ ($< T_{\rm C}$). 
The cross at $T_{\rm C}/J_1 = 0.265 \pm 0.010$ is a usual one that is found 
in the ferromagnetic phase transition\cite{Comm_Error}. 
That is, for $T > T_{\rm C}$, $g_L$ for a larger size 
is smaller than that for a smaller size; for $T < T_{\rm C}$, 
this size dependence in $g_L$ is reversed. 
On the other hand, the cross at $T_{\rm R}$ is strange. 
For $T < T_{\rm R}$, $g_L$ for a larger size again becomes smaller than 
that for a smaller size. 
Interestingly, the cross for different $g_L$ occurs at almost 
the same temperature of $T_{\rm R}/J_1 = 0.125 \pm 0.005$\cite{Comm_Error}. 
These facts suggest that, as the temperature is decreased beyond $T_{\rm R}$, 
the ferromagnetic phase, which occurs below $T_{\rm C}$, disappears.

Does the spin correlation function below $T_{\rm R}$ decay rapidly 
as the distance between two spins increases? 
If it decays according to the power law, then 
$[\langle {M(s)}^2\rangle]/xN \propto 
L^{2-\eta}$ with $\eta$ being the decay exponent of the spin 
correlation function.  
We plot, in Fig. 4, $[\langle { M(s)}^2\rangle]/xN$ as a function of $L$ 
in a log-log form at various temperatures. 
In fact, at a temperature near below $T_{\rm R}$, data seem to lie on a 
straight line; the spin correlation function will decay algebraically. 
Different $L$ dependences are found above and below this temperature. 
At higher temperatures ($T > T_{\rm R}$), as $L$ is increased, data deviate 
upward from the straight line being compatible with the fact that the 
ferromagnetic long range order occurs at these temperatures. 
On the other hand, data deviate downward at lower temperatures.  
That is, the spin correlation function will decay exponentially, 
similarly to that in the SG phase.

\begin{figure}[tb]
\begin{center}
\includegraphics[width=7.5cm,clip]{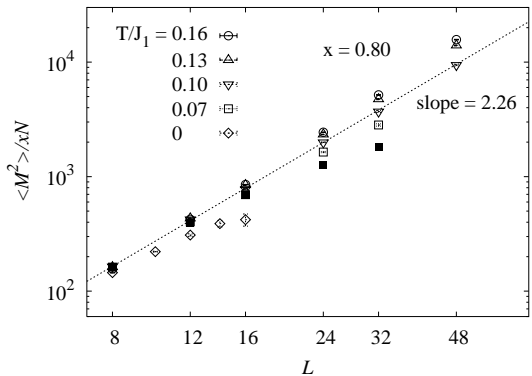}\\
\end{center}
\vspace{-0.4cm}
\caption{
\label{fig:4} 
The square of the magnetization $[\langle { M(s)}^2\rangle]/xN$ vs. $L$.  
The straight line at $T = 0.1J_1$ is obtained in the least square fitting for 
smaller lattices of $L \leq 16$. 
For $T = 0.07J_1$, open symbols indicate those in the cooling simulation and 
filled symbols represent those in the heating simulation. 
Data in both simulations deviate downward from the straight line. 
}
\vspace{-0.2cm}
\end{figure}

Is the SG phase realized at low temperatures? 
A convincing way of examining the SG phase transition is a finite 
size scaling analysis of the correlation length, $\xi_L$, in samples of 
different sizes $L$\cite{Ballesteros, Lee}. 
Data for the dimensionless ratio $\xi_L/L$ are expected to intersect at 
$T = T_{SG}$. 
Here we consider the correlation length of the SG component of the spin, i.e., 
$\tilde{\bm S}_i (\equiv {\bm S}_i - {\bm m})$ with ${\bm m}$ being the 
ferromagnetic component of ${\bm m} = \sum_ix_i{\bm S}_i/(xN)$. 
We performed a cooling simulation of a two-replica system with 
$\{{\bm S}_i\}$ and $\{{\bm T}_i\}$\cite{Bhatt}. 
The SG order parameter, generalized to wave vector ${\bm k}$, 
$q^{\mu\nu}(\bm{k})$, is defined as
\begin{eqnarray}
 q^{\mu\nu}(\bm{k}) = 
\frac{1}{xN}\sum_{i}\tilde{S}_i^{\mu}\tilde{T}_i^{\nu}e^{i{\bm k}{\bm R}_i}, 
\end{eqnarray}
where $\mu, \nu = x, y, z$. From this, we determine the wave vector dependent 
SG susceptibility $\chi_{\rm SG}({\bm k})$ by
\begin{eqnarray}
 \chi_{\rm SG}({\bm k})=xN\sum_{\mu,\nu}
                      [\langle|q^{\mu\nu}(\bm{k})|^2\rangle]. 
\end{eqnarray}
The SG correlation length is then determined from
\begin{eqnarray}
 \xi_L = \frac{1}{2\sin(k_{\rm min}/2)}
         (\frac{\chi_{SG}(0)}{\chi_{SG}({\bm k}_{\rm min})} - 1)^{1/2},
\end{eqnarray}
where ${\bm k}_{\rm min} = (2\pi/L,0,0)$.

Figure 5 shows the temperature dependence of $\xi_L/L$ for various $L$. 
In fact, $\xi_L/L$ for different $L$ intersect at $T \sim 0.10J_1$. 
In particular, data for $L \geq 12$ are scalable on the assumption that 
$T_{\rm SG}/J_1 = 0.105 \pm 0.003$. 
On the basis of this fact together with the rapid decay of the 
spin correlation function, we inferred that the SG phase is realized 
at low temperatures. 
The SG transition temperature $T_{\rm SG}$ estimated here is slightly lower 
than  $T_{\rm R}$\cite{Comm_TR}. 
However, the possibility of $T_{\rm R} = T_{\rm SG}$ cannot be ruled out, 
because the treated lattices of $L \leq 20$ for estimating $T_{\rm SG}$ 
are not large enough.

\begin{figure}[tb]
\begin{center}
\includegraphics[width=7.5cm,clip]{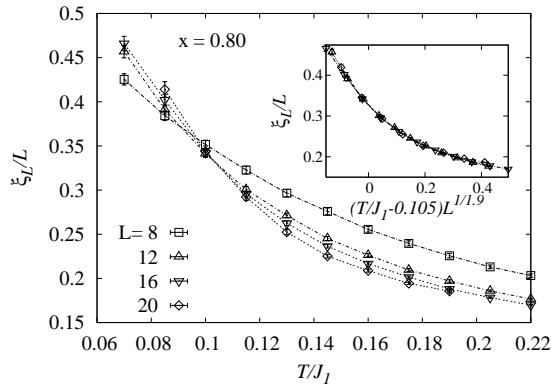}\\
\end{center}
\vspace{-0.4cm}
\caption{
\label{fig:5} 
The SG correlation length $\xi_L$ divided by $L$. The data intersect at 
$T \sim 0.10J_1$, implying that there is a SG transition at this temperature. 
The inset shows a typical example of the scaling plot for $L \geq 12$. 
}
\vspace{-0.4cm}
\end{figure}

We considered the spin structure. 
Figures 6(a) and 6(b) show typical results for it. 
In the ferromagnetic phase (Fig. 6(a)), although the spin arrangement is 
considerably modulated, a ferromagnetic spin correlation extends over 
the lattice. 
On the other hand, in the SG phase (Fig. 6(b)), 
we can see that the system breaks up to yield ferromagnetic clusters 
with a linear size of $l_c \sim 7$\cite{Comm_Corr}. 
This result is compatible with the size dependence of $M$ shown in Fig. 2. 
In the small lattice with $L = 8$, $M$ does not exhibit a marked decrease 
at $T < T_{\rm SG}$. 
The decrease becomes drastic when $L$ is increased. 
This fact is further evidence that the system is divided into 
ferromagnetic clusters. 
We suggest that the SG phase of the model is characterized by 
ferromagnetic clusters.  This concept is compatible with experimental 
observations\cite{Maletta,Aeppli,Motoya}.

In summary, we found a model that settles the most important issue 
of the RSG transition. 
Other important issues remain unresolved. 
Why does the ferromagnetic phase disappear at low temperatures? 
Does this dilute model exhibit the same behavior as that found 
in the bond SG model\cite{Matsubara,Lee}? 
Does the chiral glass phase transition simultaneously occur 
at $T_{\rm SG}$\cite{Chiral}?
We intend the present model as one means to solve those and other 
remaining problems.

\begin{figure*}[tb]
\hspace{-1.2cm} (a) $T = 0.19J_1$
\hspace{4.7cm}  (b) $T = 0.04J_1$\\
\vspace{0.2cm}
\hspace{1cm}\includegraphics[width=5.0cm,clip]{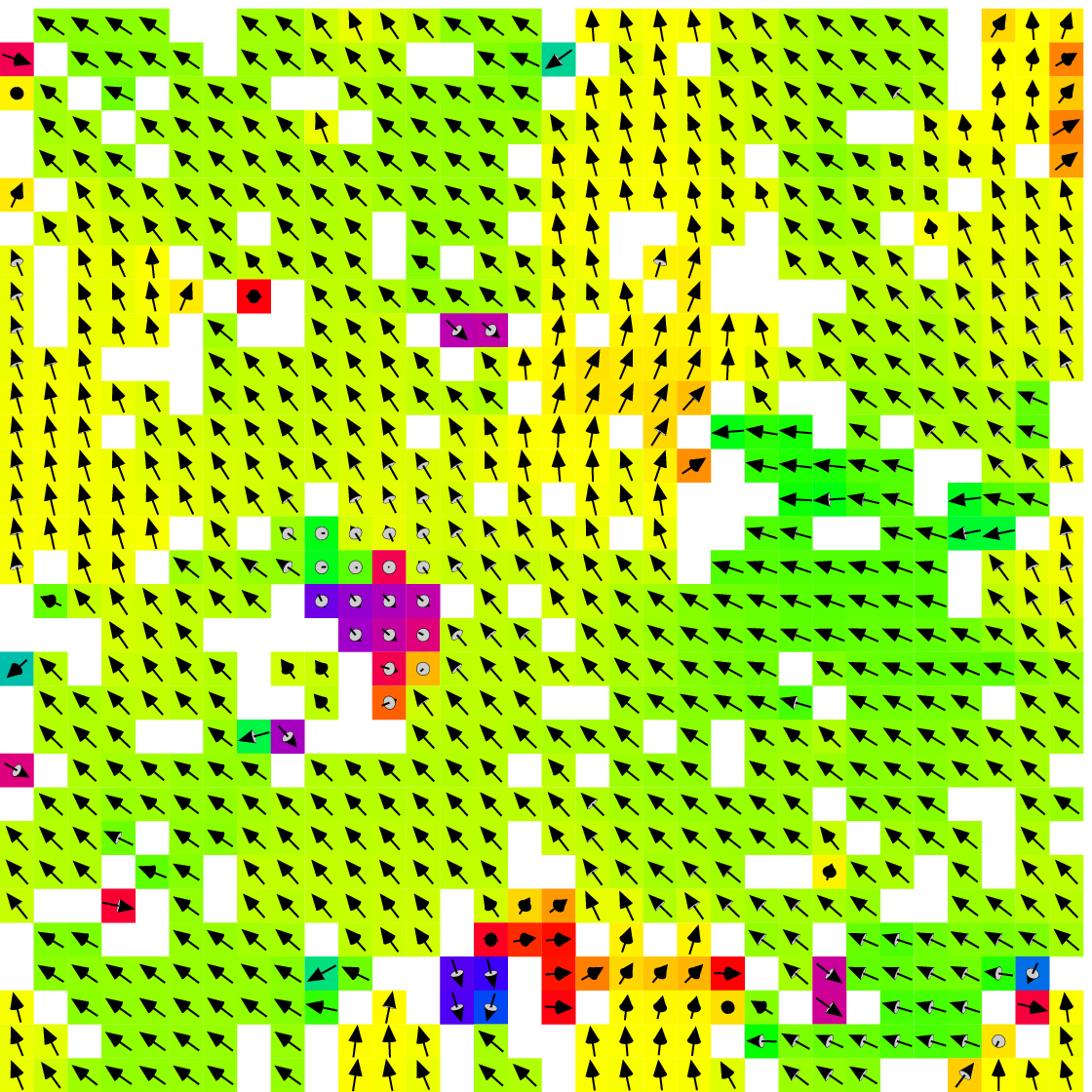}
\hspace{2cm}\includegraphics[width=5.0cm,clip]{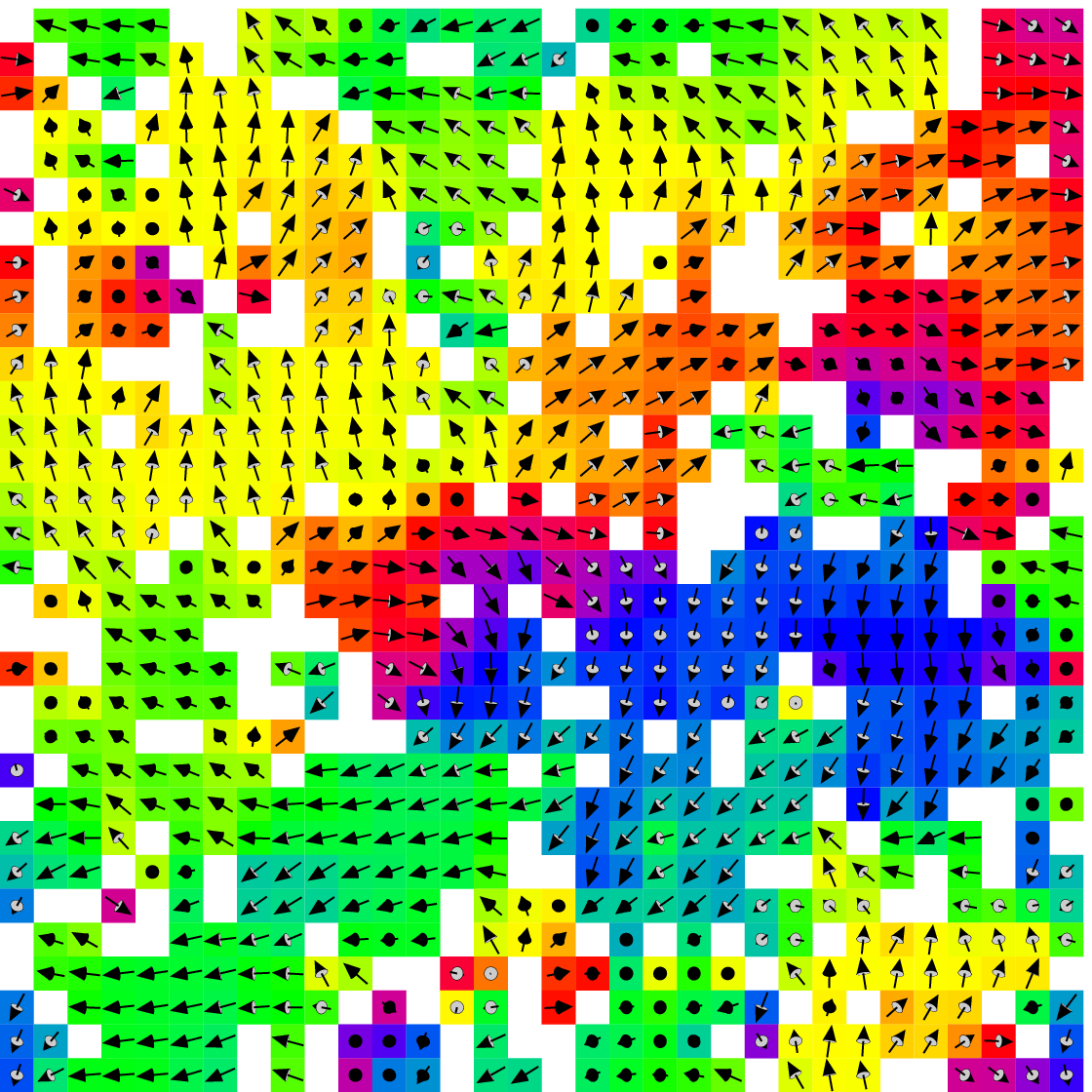}
\hspace{0.5cm}\includegraphics[width=1.5cm]{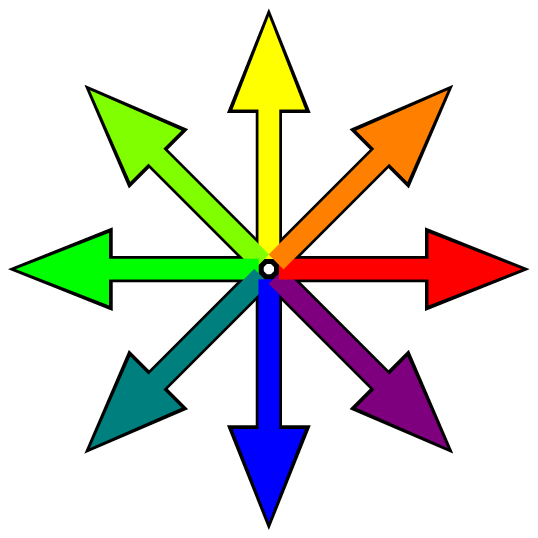}
\caption{ \label{fig:6}
Spin structures of the model for $x = 0.80$ on a plane of the 
$32 \times 32 \times 32$ lattice in (a) the ferromagnetic phase 
and (b) the spin glass phase. Spins represented here are those averaged 
over 10000 MCS.  
The positions of the non-magnetic atoms are represented in the white.
}
\end{figure*}

\bigskip
\bigskip

The authors are indebted to Professor K. Motoya for directing their attention 
to this problem of the RSG transition and for his valuable discussions. 
The authors would like to thank Professor T. Shirakura for his useful 
suggestions and Professor K. Sasaki for critical reading of the manuscript. 
This work was financed by a Grant-in-Aid for Scientific Research 
from Ministry of Education, Science and Culture.



\begin{references}
\bibitem{Model} Coordination numbers of the nearest and the next-nearest 
          neighbor lattice sites are $z_1 = 12$ and $z_2 = 6$ in the fcc 
          lattice, and $z_1 = 6$ and $z_2 = 12$ in the sc lattice. 
          In that case, we choose the smaller ratio of $J_2/J_1 = 0.2$ 
          instead of $J_2/J_1 \sim 0.5$ in Eu$_x$Sr$_{1-x}$S.
\bibitem{Comm_gL} 
         Values of $g_L$ depend strongly on the sample number $N_s$, 
         especially at low temperatures ($T \lesssim 0.15 J_1$). 
         Figure 3 shows $g_L$ only when the two restrictions are 
         fulfilled: (i) $g_L$s for both cooling and heating simulations 
         mutually coincide; and (ii) $g_L$ for $N_s$ is not 
         changed considerably from that for $N_s/2$.
\bibitem{Comm_TR} This is possible if the ferromagnetic order parameter 
         ${\bm m}$ and the SG order parameter decouple from each other. 
         The difference in the transition temperatures of 
         $T_{\rm INV}( \equiv T_{\rm R})$ and $T_{\rm SG}$ was reported 
         in the magnetic alloy of Fe$_{0.7}$Al$_{0.3}$\cite{Motoya}.
\bibitem{Comm_Corr} 
         We have estimated the correlation length of $\xi \sim 3.5$ 
         using the function of $\langle {\bm S}_0{\bm S}_{r}\rangle = 
         a\exp(-r/\xi)/r^{1+\eta} + m^2$ with $a$ and $m$ being 
         constants; the domain size of $l_c \sim 2\xi \sim 7$.
\end{references}
\end{document}